**Manuscript title:** A tutorial on evaluating time-varying discrimination accuracy for survival models used in dynamic decision-making

**Running head:** Biomarker-driven dynamic decision-making

**Authors:** Aasthaa Bansal, PhD[1*], Patrick J. Heagerty, PhD[2]

[1] The Comparative Health Outcomes, Policy, and Economics Institute, School of Pharmacy, University of Washington, Seattle WA, USA
Email: abansal@uw.edu
Phone: (206) 427-5448

[2] Department of Biostatistics, University of Washington, Seattle WA, USA
Email: heagerty@uw.edu
Phone: (206) 616-2720

[*] Corresponding author
Address: The Comparative Health Outcomes, Policy, and Economics Institute
School of Pharmacy
University of Washington
H-375, Health Sciences Building
Box 357630
Seattle WA, USA, 98195

**Department(s) and institution(s) where work was done:**
- Department of Biostatistics, University of Washington, Seattle WA, USA
- The Comparative Health Outcomes, Policy, and Economics Institute, School of Pharmacy, University of Washington, Seattle WA, USA

**Meetings at which the work was presented:**
- Society for Medical Decision Making annual meeting, St. Louis, MO, 2015
- Joint Statistical Meetings, American Statistical Association, Seattle, WA, 2015

**Grant support:** This research was supported by the PhRMA Foundation, the National Heart, Lung, and Blood Institute of the National Institutes of Health (NIH) (under R01-HL072966), and by the National Center For Advancing Translational Sciences of the NIH (under UL1TR000423).

**Word count (excluding title page, abstract, acknowledgments, references, and figure legends):** 4,999

*Financial support for this study was provided in part by grants from the PhRMA Foundation, the National Heart, Lung, and Blood Institute of the National Institutes of Health (NIH), and by the National Center For Advancing Translational Sciences of the NIH. The funding agreement ensured the authors' independence in designing the study, interpreting the data, writing, and publishing the report.*

**ABSTRACT  (Word count: 254 / 275)**

Many medical decisions involve the use of dynamic information collected on individual patients toward predicting likely transitions in their future health status. If accurate predictions are developed, then a prognostic model can identify patients at greatest risk for future adverse events, and may be used clinically to define populations appropriate for targeted intervention. In practice, a prognostic model is often used to guide decisions at multiple time points over the course of disease, and classification performance, i.e. sensitivity and specificity, for distinguishing high-risk versus low-risk individuals may vary over time as an individual's disease status and prognostic information change. In this tutorial, we detail contemporary statistical methods that can characterize the time-varying accuracy of prognostic survival models when used for dynamic decision-making. Although statistical methods for evaluating prognostic models with simple binary outcomes are well established, methods appropriate for survival outcomes are less well known and require time-dependent extensions of sensitivity and specificity to fully characterize longitudinal biomarkers or models.   The methods we review are particularly important in that they allow for appropriate handling of censored outcomes commonly encountered with event-time data. We highlight the importance of determining whether clinical interest is in predicting cumulative (or prevalent) cases over a fixed future time interval versus predicting incident cases over a range of follow-up times, and whether patient information is static or updated over time. We discuss implementation of time-dependent ROC approaches using relevant R statistical software packages. The statistical summaries are illustrated using a liver prognostic model to guide transplantation in primary biliary cirrhosis.

# 1 Introduction

Many medical decisions involve using updated information on patients under surveillance to predict transitions in future health status, such as progression of disease or advancement to death. The goal is to use a patient's clinical characteristics to calculate the predicted risk of an event within a specified time period and to identify patients who are at high risk of experiencing an adverse event in the near future. If accurate predictions can be made, they could be used clinically to guide the choice and timing of interventions and enable timely action, such as starting specific preventive strategies or initiating aggressive treatments for high-risk individuals while sparing low-risk patients the side-effects and costs of unnecessary intervention.

In practice, prognostic models are often used to make decisions at multiple time points over the course of patient follow-up. Consider disease screening settings, where predicted risk may be used to identify high-risk individuals as candidates for more frequent screening. Patient follow-up with updated clinical assessment also frequently occurs to monitor response to therapy. For example, a cancer patient who has previously undergone treatment and is predicted to be at substantial risk of disease recurrence may benefit from adjuvant therapy, whereas a low-risk patient may forego aggressive treatment. Finally, in an organ transplantation setting, the predicted risk of mortality may be used to guide prioritization and timing of donor organ transplantation.[1-3]

Traditional statistical models such as Cox regression focus on the prediction of disease or death times. However, underlying these standard methods are the concepts of a time-varying "risk set" of individuals, and associated time-specific "cases" or subjects who experience the clinical event (ie. death) at a given time. At any time point, the set of individuals still alive and at risk of an event may be partitioned into imminent "cases" (individuals who experience the event in a defined future time frame) and current "controls" (individuals who do not yet experience the event). Ultimately, the goal of a prognostic model is to accurately predict event times, or equivalently to distinguish between the time-specific cases and the controls at all follow-up times. Furthermore, in practice an individual's disease status changes over time, and so does his or her prognostic information, such as laboratory measures updated in routine clinic visits. Accordingly, a model's ability to distinguish between cases and controls over time may also change, thus impacting its performance as a decision-making tool. For example, a prognostic model may accurately identify patients at high risk of death within 90 days, but it may have reduced accuracy for identifying later deaths.

Accuracy concepts of sensitivity and specificity are fundamental to clinical research and decision modeling. Only recently have statistical methods been developed that can generalize these traditionally cross-sectional accuracy concepts for application to the time-varying nature of disease states, and corresponding definitions of time-dependent sensitivity and specificity have been proposed for both prevalent and incident case definitions.[2,3] These new concepts and associated statistical methods are central to the evaluation of the time-varying performance of any potential prognostic model; they allow for the estimation of sensitivity, specificity and area under the receiver operating characteristic (ROC) curve (AUC) as functions of time, thus providing a detailed estimate of longitudinal model performance for use in practice. These methods are particularly important in that they allow for appropriate handling of right-censored outcomes commonly encountered with clinical event time data. Unfortunately, knowledge of these methods and the tools available to implement them remains limited, and investigators often resort to overly simplistic application of methods developed for binary outcomes, which can lead to biased estimates in the presence of censoring.[4-5]

Our goal in this tutorial is to demonstrate the use of modern statistical methods that address the following questions: how can the time-varying discrimination accuracy of a prognostic model be evaluated; how can the value of updated measurements be characterized; and how can different candidate models be directly compared? We highlight the importance of determining whether interest is in the fundamental epidemiologic concept of predicting cumulative (or prevalent) cases, or in incident cases.

## 1.1 Case study: Liver Prognostic Model to Guide Transplantation in Primary Biliary Cirrhosis

As an illustrative case study, we consider liver transplantation in primary biliary cirrhosis (PBC). PBC is an autoimmune disease in which the bile ducts are slowly destroyed, leading to liver failure in cases of advanced disease.[6] For selected patients with liver failure who are at high risk of death without transplantation, liver transplantation can be potentially life-saving. As a result, a number of prognostic models have been developed in PBC, with the goal of predicting survival probabilities and guiding decisions regarding transplantation.[7-13] Of these, the Mayo model is perhaps the most widely known[7] with the more recent Model for End-stage Liver Disease (MELD) score[2] representing a refinement, but potentially suboptimal for use in PBC.[7] A unique characteristic of the Mayo model compared to other existing models is that it does not require liver biopsy. Instead, it is based on inexpensive, noninvasive and readily available measurements. Additional variables from a biopsy, such as histologic stage, that are used in other models have been shown to not contribute substantially beyond the variables included in the Mayo model.[1]

We consider a well-known dataset that comes from a randomized placebo-controlled trial for the treatment of PBC conducted at the Mayo Clinic between 1974 and 1984.[14] Dickson et al.[1] used this data to develop the Mayo risk model that included patient age, total serum bilirubin and serum albumin concentrations, prothrombin time, and severity of edema. This model has been used for making individual-level decisions regarding the selection of patients for and timing of liver transplantation in PBC.[7] Decisions about transplantation are made repeatedly over time, by selecting patients who are most likely to die in a short time interval, such as 90 days, 6 months or 1 year from the time of prediction. We will use the five main predictors of survival identified by Dickson et al.[1] to calculate the predicted risk of mortality within specified time periods, and evaluate the accuracy of these predictions for prioritizing patients for transplantation.

## 1.2 Model Development

Model development typically takes place by splitting a dataset into training and validation data that are used for model selection and evaluation, respectively. Using appropriate methods to avoid overfitting in the training data,[15-17] candidate biomarkers and variables are selected and combined, traditionally using a Cox proportional hazards regression model for survival outcomes.[18] One may use standard Cox regression with fixed coefficients and baseline covariates, or even incorporate time-varying covariates, as well as time-varying coefficients into the model.[19] Alternatively, one may use more flexible, modern machine-learning approaches, such as boosting, lasso, artificial neural networks, and random forests, especially in the presence of high-dimensional data.[20-26] Regardless of the chosen modeling approach, the ultimate prognostic model is then fixed and used in the validation data to provide patient predictions of the disease outcome, i.e. a risk score.

In this manuscript, we are agnostic to model selection. We focus on methods for evaluating any single "biomarker", which may be a novel predictive measurement, such as a specific serum protein level measured in the laboratory, or more commonly may be the risk score derived from a model that includes multiple factors, i.e. a *derived* biomarker or classifier. The approaches we discuss for evaluating a risk score in the validation data are independent of those used for model selection in the training data, in that they do not rely on the assumptions that may be necessary for the development of the risk score.

Given our focus on model evaluation, it is not our objective here to develop a new model as an alternative to the Mayo model. We simply demonstrate how to evaluate the time-varying performance of the existing Mayo risk score, as well as one variation of it where we omit a variable, in order to demonstrate a comparison of two candidate models.

## 2 Background: Standard Measures of Discrimination Accuracy

The traditional classification problem is based on a simple binary outcome, typically the presence or absence of disease. In classifying cases and controls as having disease or not, a marker is prone to two types of error: incorrectly classifying a case as not having disease, leading to delays in treatment, and conversely, incorrectly classifying a control as having disease, subjecting the individual to unnecessary follow-up medical procedures. Investigators aim to minimize false negative and false positive errors by developing markers with high sensitivity (true positive fraction (TPF)) and high specificity (1 minus false positive fraction (FPF)), respectively.

By convention, larger marker values are assumed to be more indicative of disease (and if the opposite is true, the marker is transformed to fit the convention). For a continuous marker $M$ and a fixed threshold $c$, we define

$$\text{sensitivity}(c) = \text{P}(M > c \mid \text{case}),$$
$$\text{specificity}(c) = \text{P}(M \leq c \mid \text{control}).$$

The Receiver Operating Characteristic (ROC) curve is a standard tool that plots a continuous marker's sensitivity against 1-specificity for all possible values of the threshold $c$.[27-30] Classification accuracy is most commonly summarized using the area under the ROC curve (AUC), which is the probability that a randomly chosen case has a higher marker value than a randomly chosen control:

$$\text{AUC} = P(M_i > M_j \mid i = \text{case}, j = \text{control}).$$

Therefore, the AUC represents the marker's ability to rank cases above controls. An AUC of 0.5 indicates no discrimination between cases and controls, whereas an AUC of 1.0 indicates perfect discrimination.[30]

## 3 Time-Dependent Discrimination Accuracy

Implicit in the use of traditional diagnostic sensitivity and specificity are current-status definitions of disease. In settings of long-term follow-up, disease status changes with time and precise definitions are necessary to include event (disease) timing in definitions of prognostic error rates. Within the last two decades, time-dependent ROC curve methods that extend concepts of sensitivity and specificity and characterize prognostic accuracy for survival outcomes have been proposed in the statistical literature and adopted in practice. We review two such time-dependent approaches, which draw upon alternative fundamental case definitions: cumulative (or prevalent) cases; and incident cases.

### 3.1 Cumulative (Prevalent) Cases / Dynamic Controls

Often interest lies in identifying individuals at risk of an adverse event within some fixed time frame. Recall, for example, decisions about donor liver allocation in the PBC setting being made by selecting patients who are most likely to die in a short time interval, such as 90 days, 6 months or 1 year, from the time of prediction.

A natural extension of the standard cross-sectional definitions of sensitivity and specificity to the survival context, where disease state is time-dependent, is to dichotomize the outcome at a selected time of interest, $t$ (90 days, 6 months or 1 year), and define cases as subjects who experience the event before time $t$, and controls as those who remain event-free beyond $t$.[31] More formally, we let $T$ denote survival time and $s$ denote the start time of case ascertainment (often $s=0$ for baseline). Then, cumulative cases ($C$) may be defined as subjects who experience an event prior to $t$, or specifically as $T_i \, \epsilon \, (s,t)$, and dynamic controls ($D$) as subjects who are event-free at time $t$, $T_i > t$ (regardless of whether or not they experience the event at a later time). Then for a fixed threshold $c$, time-dependent definitions for sensitivity and specificity follow[31,32]:

$$\text{sensitivity}^{\text{C}}(c \mid \text{start} = s, \text{stop} = t) = \text{P}(M > c \mid T \geq s, \, T \leq t)$$

$$\text{specificity}^{\text{D}}(c \mid \text{start} = s, \text{stop} = t) = \text{P}(M \leq c \mid T \geq s, \, T > t)$$

Let $p$ represent a fixed FPF. Then, for fixed specificity$^{\text{D}}(c|s,t) = 1 - p$, the time-dependent ROC value is the corresponding value of sensitivity$^{\text{C}}(c|s,t)$, or $\text{ROC}_{s,t}^{\text{C/D}}(p)$. Correspondingly, the time-specific AUC is defined as the area under the time-specific ROC curve across all thresholds $p$:

$$\text{AUC}^{\text{C/D}}(s, t) = \int \text{ROC}_{s,t}^{\text{C/D}}(p) \, dp$$

which can be shown to be equivalent to

$$\text{AUC}^{\text{C/D}}(s, t) = \text{P}(M_j > M_k \mid T_j \geq s, \, T_j \leq t, \, T_k \geq s, \, T_k > t).$$

Here, $\text{AUC}^{\text{C/D}}(s,t)$ is the probability that a random subject $j$ who experiences an event before time $t$ (case) has a larger marker value than a random subject $k$ who remains event-free through time $t$ (control), assuming both subjects are event-free at the start of follow-up, time $s$.

In the absence of censoring, the above dichotomization at time $t$ is equivalent to using a simple derived binary disease outcome. However when follow-up is incomplete, as is often the case with longitudinal data, censoring needs to be addressed and can be handled using nonparametric estimation of the bivariate distribution of $(M,T)$.[31] (See Appendix A for description of estimation methods). Estimation is based on $(Z_i, \delta_i)$, where $Z_i$ is the observed follow-up time, i.e. the minimum of the survival time $T_i$ and the right censoring time $C_i$, and $\delta_i$ denotes the event indicator.

In this tutorial, we seek to characterize time-varying performance over a meaningful range of times. To this end, we suggest obtaining a sequence of accuracy assessments over time by defining cases as events occurring cumulatively in successive windows of time. Specifically, we subset data at a sequence of index times $s = t_1, t_2, ..., t_K$ to include only subjects who are event-free at time $t_k$, i.e. $Z \geq t_k$, $k=1,...,K$. These index times can represent any time points of interest and do not have to fall at constant time intervals. For each subsetted dataset, we suggest conducting a separate analysis, treating $t_k$, $k$=1,...,$K$, as the new baseline $s$ and defining cases cumulatively as subjects who have events over the following, say, 1-year span, so that $Z_i \in (s = t_k, t = t_k + 1)$ and $\delta_i$=1, and defining controls such that $Z_i > t_k + 1$ (Figure 1). A series of accuracy summaries, such as AUC[C/D](0, 1), AUC[C/D](2, 3), AUC[C/D](4, 5), ..., is obtained, and time-varying accuracy is indicated by a change in AUCs over time. The same idea can be applied to obtain time-varying sensitivity and specificity.

If prognostic information changes over time, updated information can be included in each subsetted analysis by using the last measured information to obtain updated risk predictions. Although we chose a 1-year cumulative window for illustration, the window is flexible and may be chosen to be more clinically meaningful depending on the disease setting. Alternatively, the incident/dynamic approach, discussed next, provides a finer timescale, allowing for a smoother characterization of performance over time without having to specify a window of time over which cases accumulate.

## 3.2 Incident Cases / Dynamic Controls

Survival analysis using Cox regression is based on the fundamental concept of a risk set: a risk set at time $t$ consists of the cases experiencing events *at* time $t$, and the additional individuals who are under study (alive) but do not yet experience the clinical event. Extension of binary classification error concepts to risk sets leads naturally to adopting an incident (*I*) case definition where subjects who experience an event *at* time $t$ or have survival time $T_i = t$ are the time-specific cases of interest. Dynamic controls (*D*) can be compared to incident cases and are subjects with $T_i > t$ (regardless of whether or not they experience the event or get censored at a later time). In this scenario, time-dependent definitions for sensitivity and specificity are[33]:

$$\text{sensitivity}^I(c \mid t) = P(M > c \mid T = t)$$
$$\text{specificity}^D(c \mid t) = P(M \leq c \mid T > t)$$

For fixed specificity$^D(c|t) = 1 - p$, the time-dependent ROC value is the corresponding value of sensitivity$^I(c|t)$, or $\text{ROC}_t^{I/D}(p)$. The time-dependent AUC can be defined as the area under the time-specific ROC curve across all thresholds $p$:

$$\text{AUC}^{I/D}(t) = \int \text{ROC}_t^{I/D}(p) \, dp$$

which can be shown to be equivalent to

$$\text{AUC}^{I/D}(t) = P(M_j > M_k \mid T_j = t, T_k > t).$$

Here, $\text{AUC}^{I/D}(t)$ is the probability that a random subject $j$ who experiences an event at time $t$ (case) has a larger marker value than a random subject $k$ who remains event-free through time $t$ (control), assuming both subjects are event-free up to time $t$.

A semiparametric method based on the Cox model[33], as well as a nonparametric rank-based method[34], have been proposed for estimating $\text{ROC}_t^{I/D}(p)$ and $\text{AUC}^{I/D}(t)$ with censored outcomes. Both methods estimate $\text{FPF}_t^D$ nonparametrically; the difference comes from their estimation of $\text{TPF}_t^I$, which requires smoothing since the observed subset with $T_i = t$ may only contain one observation. The semiparametric method achieves smoothing by fitting a hazard model, whereas the nonparametric method uses kernel-based smoothing (See Appendix A for additional details). The nonparametric approach is generally preferable as it relies on fewer assumptions than the semiparametric approach. Additionally, the nonparametric method has been developed to provide a simple summary curve that graphically characterizes accuracy over time.

Furthermore, the performance of updated prognostic information can also be evaluated by using the semiparametric[33] or nonparametric[34] approach to accommodate time-varying markers.[35] At any time $t$, the last measured information may be used to obtain updated risk predictions from the prognostic model, as discussed in the previous section.

### 3.2.1 Global Summary of Marker Performance

In many applications, there is no specific time $t$ of interest, and a global accuracy summary of time-varying performance is desired. Furthermore, it may also be of interest to compare the overall performance of different markers or models. The incident/dynamic approach lends itself easily to addressing such questions, since marker performance can be summarized into a single-number global summary called the survival concordance index (c-index)[33]:

$$\text{c-index} = P(M_j > M_k \mid T_j < T_k).$$

The c-index is interpreted as the probability that the predictions for a random pair of subjects are concordant with their outcomes. In other words, it is the probability that the subject who died at an earlier time had a larger marker value. The c-index can also be expressed as a weighted average of time-specific AUCs[33] and is therefore easy to estimate using the incident/dynamic methods described above. The above definition of the basic c-index for survival outcomes applies to a baseline marker $M$. However, the definition and associated estimation methods can easily be generalized to accommodate updated prognostic information to estimate the generalized c-index for a time-varying marker, $M(t)$, expressed as:

$$\text{generalized c-index} = \int \text{AUC}^{I/D}(t) \, w(t) \, dt$$

using the weighted average representation which allows time-varying markers to be use for each $\text{AUC}^{I/D}(t)$ (See Appendix A for definition of w(t) with further details, and Section 4 for an illustration).

### 3.3 Extension to competing risk outcomes

Often a subject's event time can be classified by one of several distinct causes and interest may lie in events of a specific type. For example, in breast cancer studies, distant metastasis may be the event of interest; however, other clinical events, such as death, may preclude the researcher from observing distant metastases for particular patients.[36] The definitions of time-dependent sensitivity, specificity, ROC and AUC presented in Sections 3.1 and 3.2 have been extended to incorporate cause of failure for competing risk outcomes for both the cumulative and incident case definitions and we direct the reader to the associated literature.[37]

### 3.4 Software

The above methods have been implemented in publicly available `R` statistical software packages `survivalROC` (for cumulative/dynamic methods), `risksetROC` (for incident/dynamic methods with semiparametric estimation) and `meanrankROC` (for incident/dynamic methods with nonparametric estimation). The cumulative/dynamic methods have also been implemented as part of the `PHREG` procedure in the commercial software SAS. These software options are summarized in Table 1. Additionally, the `survivalROC` and `risksetROC` packages have been extended to include updated definitions for competing risk outcomes.

We note that the choice of `R` package should depend on the chosen method, which should depend on the scientific question of interest, as discussed in Section 3.5 and illustrated using the `survivalROC` and `meanrankROC` packages in the case study of Section 4 (with accompanying code in Appendix B).

### 3.5 Comparison of Cumulative versus Incident Case Approaches

Use of incident events naturally facilitates evaluation of time-varying prognostic performance, whereas the use of cumulative events in a sequential manner can also enable such evaluation. In practice, patterns in $\text{AUC}^{I/D}(t)$ tend to match $\text{AUC}^{C/D}(t,t+1)$ closely when the gap between $t$ and $t+1$ is small, although $\text{AUC}^{C/D}(t,t+1)$ uses a coarser time scale and averages the performance over a fixed time interval.

In a descriptive context, $\text{AUC}^{I/D}$ may be preferable because it provides a simple graphical approach and a global summary using the c-index, without having to specify a time interval over which cases accumulate. In contrast, sequential use of cumulative cases based on $\text{AUC}^{C/D}$ may better align with clinical settings where prediction of short-term survival is needed at a specific decision time (or a small collection of times). For example, time intervals of 6 months, 1 year and 5 years are commonly used for defining high-risk versus low-risk patients for targeted intervention. Methods for meaningfully averaging time-varying performance into a global performance summary using the cumulative case definition have not been developed.

Computationally, $\text{AUC}^{I/D}(t)$ is more straightforward to estimate and visualize for a series of time points. $\text{AUC}^{C/D}(t)$ requires the generation of a new subsetted dataset for each time point of interest and therefore if interest lies in several time points, then a series of $\text{AUC}^{\text{C/D}}(t)$ estimates may be more cumbersome to obtain.

# 4 Case study: Liver Prognostic Model to Guide Transplantation in Primary Biliary Cirrhosis

## 4.1 Description of Study Cohort

The study cohort consisted of 312 patients; 125 (40%) of these patients were observed to die during the study period; 19 subjects were recipients of liver transplantation during the study period. We censored these subjects at the time of transplantation, since the prognostic model is intended to predict the risk of mortality *without transplantation* and use that risk to prioritize such patients. For each patient, we had baseline demographic and diagnosis data and longitudinal data on laboratory measures. Counting multiple observations per patient, we included 1,945 total records.

## 4.2 Risk models

We evaluated the following models: (i) a 5-covariate model containing the same variables as those in the Mayo model[1]: log(bilirubin), albumin, log(prothrombin time), edema and age, and (ii) a 4-covariate model where we omitted log(bilirubin) to illustrate the comparison of different candidate models. Predictions from Cox models were summarized into a single baseline risk score and a separate time-varying, updated risk score, in order to demonstrate that the methods can incorporate time-varying measurements and to show the implications of using older measurements on accuracy. For the baseline score, we used 10-fold cross-validation to protect against overfitting.[15-17] For the time-varying score, we used baseline measurements as training data to develop the Cox model and predicted the score at follow-up times using updated values of log(bilirurbin), albumin, and log(prothrombin time).[15-17]

## 4.3 What is the accuracy of baseline measurements and the value of updated measurements?

As a first step, we use the incident/dynamic approach to assess the prognostic accuracy of the baseline risk score obtained from the 4-covariate model versus the 5-covariate model. Figure 2 and Table 3 show that the 5-covariate model has consistently better performance than the 4-covariate model over time with respect to both $AUC^{I/D}(t)$ (Table 3 and Figure 2, left panel) and sensitivity for a fixed specificity of 10% (Figure 2, right panel). The estimated c-indices are 0.72 (95% CI: (0.66, 0.76)) and 0.79 (95% CI: (0.75, 0.83)) for the 4- and 5-covariate models, respectively, with a statistically significant difference of 0.07 (95% CI: (0.04, 0.11)). Table 3 also shows the sequential cumulative/dynamic approach that uses successive 1-year windows to define cases. We see similar estimates for $AUC^{I/D}$ and $AUC^{C/D}$. Any observed differences are due to $AUC^{I/D}$ reflecting performance at a given time point and $AUC^{C/D}$ averaging performance over a 1-year window.

Looking at the 5-covariate model, the performance of the baseline score declines over time with $AUC^{I/D}$ = 0.88 (95% CI: (80, 0.90)) at 1 year versus 0.66 (95% CI: (0.62, 0.78)) at 6 years. In contrast, fairly consistent performance is maintained using a risk score that is updated over time ($AUC^{I/D}(t)$ = 0.92 (95% CI: (0.88, 0.96)) at 1 year, 0.89 (95% CI: (0.84, 0.92)) at 6 years) (Table 3 and Figure 3). 95% confidence intervals are included in Table 3, and can also be included in plots, as shown in Figure 4 for baseline and updated risk scores from the 5-covariate model.

Similar patterns are observed for the 4-covariate model, with the baseline score's performance declining over time and the updated risk score's performance staying fairly steady. Interestingly, the updated 4-covariate risk score performs almost as well as the updated 5-covariate risk score, indicating that some of the loss of accuracy due to the omission of log(bilirubin) can be recovered by using updated measurements on other variables.

## 4.4 Implications for Decision-Making in PBC

This Mayo risk score has been used for individual-level decision-making about transplantation over time, by selecting patients who are most likely to die in a short time interval from the time of prediction. We used the five main predictors of survival identified by Dickson et al.[1] to calculate the predicted risk of mortality and evaluate the accuracy of these predictions toward prioritizing patients for transplantation. It is clear from the results that patient information should be updated regularly in practice, in order to maintain prognostic accuracy. The updated 5-covariate Mayo model maintains an $AUC^{I/D}$ of around 0.90 over time, with a high generalized c-index of 0.89 (95% CI: (0.84, 0.92)), indicating that it is a strong prognostic model for use in practice. Additionally, we used $AUC^{C/D}$ sequentially with 1-year windows to evaluate the use of the Mayo model as a decision-making tool in practice. We found that $AUC^{C/D}$ is consistently above 0.80 at all chosen time points, indicating that the model identifies high-risk patients for transplantation with good accuracy.

## 5 Discussion

The American Heart Association's 2009 criteria for evaluating a risk prediction model categorize performance measures into those of calibration, association, discrimination, and risk reclassification.[38] Similarly, Steyerberg et al[39] differentiated the roles of various performance measures for assessing prediction models, defining them as measures of overall performance, discrimination, calibration, reclassification, and clinical usefulness. They explained that these measures serve different purposes and suggested that "reporting discrimination and calibration will always be important for a prediction model". Although their focus was on binary outcomes, the same ideas hold for survival outcomes.

In this tutorial, we focused on discrimination accuracy (other work has demonstrated calibration for prognostic models for survival outcomes[40]). We presented methods that extend standard diagnostic definitions of sensitivity and specificity and develop key summaries for evaluating the time-varying prognostic performance of a marker or model measured at baseline only or updated in routine clinical care. A basic epidemiologic concept that distinguishes alternative summaries is the idea of cumulative versus incident events to define cases. $AUC^{I/D}(t)$ is a convenient descriptive and graphical summary that characterizes time-varying performance without having to select a particular timeframe over which cases accrue, whereas sequential use of $AUC^{C/D}(t)$ may be useful in clinical settings where predictions of short-term survival are needed at select times to identify high-risk patients for targeted intervention.

In addition to allowing for evaluation of time-varying discrimination accuracy of prognostic models, there are other implications for how these methods could be applied in practice. First, these methods may guide practice and policy with regards to the frequency of updating patient information, by comparing the performance of risk scores updated using different measurement schedules to assess how often patient information should be updated before it becomes outdated and impacts accuracy. Second, although we compared the 5-covariate Mayo model to a simple 4-covariate variation of the model for illustration, in practice, one may choose more clinically relevant variables, such as more expensive measures, to omit or replace and assess the impact on prognostic accuracy. Finally, one may choose to explore the performance of a risk model in subsets of patients, say older versus younger patients, to assess whether the model is a better decision-making tool for particular subgroups.

One limitation of this tutorial is that we do not discuss model selection in detail, focusing on the evaluation of a given model. However, the methods for model evaluation that we discuss could also be used at the stage of model selection to guide identification of a model with optimal performance. For example, with variable selection in high-dimensional settings, one may use the c-index, which is a global summary of time-varying performance, as a way of initially screening the strongest markers as candidates for combining into a multivariate risk score. One may also use the c-index as the optimization criterion in model selection, instead of the typically used likelihood-based criteria.[41-43] For example, approaches that optimize the c-index have been developed using boosting.[44,45]

A potential limitation of the case study is that in the absence of an independent dataset on PBC, our illustration of methods for evaluation uses the same dataset that was used by Dickson et al[1] to develop the Mayo model. As discussed in Section 1.2, the standard approach is to use separate training and validation datasets to fairly assess model performance. We used cross-validation to mitigate the potential issue of an optimistic assessment. In practice, an independent validation dataset is important if the results may have clinical implications. However, this case study was meant to illustrate methods, rather than inform clinical practice. Additionally, the case study uses data from a trial conducted between 1974 and 1984. Again, a newer dataset would not add substantially to our primary goal of illustrating methods. Furthermore, the Mayo model, which is widely used in practice today, was developed using the same dataset.

Finally, there is growing interest in evaluating the incremental value gained from adding a new marker(s) to an existing baseline marker or model. Difference in AUC is a popular metric for evaluating incremental value. As we illustrated using the case study, the time-varying incremental value of a marker can be evaluated by comparing the time-varying AUCs of two models. Additionally, a number of alternative measures have been proposed in recent literature for binary outcomes, namely the net reclassification index[46] and integrated discrimination improvement[47]. Extensions of these measures for time-dependent outcomes have been developed[48,49] and can provide alternative summaries of the time-varying incremental value of a marker.

## Acknowledgements

This research was supported by the PhRMA Foundation, the National Heart, Lung, and Blood Institute of the National Institutes of Health (NIH) (under R01-HL072966), and by the National Center For Advancing Translational Sciences of the NIH (under UL1TR000423).

# Tables

Table 1: A guide to available software for conducting analyses using the cumulative/dynamic and incident/dynamic methods

| Measures of Interest | Software |
|---|---|
| **Cumulative cases/Dynamic controls** | R package `survivalROC` |
| • ROC function | `survivalROC.C()` accepts censored survival data and returns a set of TPF and FPF values for construction of the ROC curve, $ROC_{s,t}^{C/D}$, where $s$ is the "baseline" time of the subsetted dataset, i.e. $T \geq s$, while $t$ (specified using the `predict.time` argument) defines the window over which cases accumulate, so that $T \leq t$ defines cases and $T > t$ defines controls. The function calculates estimates and associated 95% confidence intervals for $ROC_{s,t}^{C/D}(p)$ on subsetted datasets based on new index (or "baseline") times and updated marker values. |
| • AUC function | `survivalROC.C()` (described above) also calculates estimates and associated 95% confidence intervals for $AUC^{C/D}(s,t)$. |
| • Example | The documentation for the `survivalROC` package demonstrates the above functionality on *baseline* markers in the Mayo PBC dataset. Furthermore, see Section 4 of this tutorial (and Appendix B for corresponding R code) for an illustration of the package applied to assessing *time-dependent* discrimination accuracy of both baseline and *time-varying* markers. |
| **Cumulative cases/Dynamic controls** | `SAS procedure PHREG` |
| • ROC function | The PHREG procedure accepts censored survival data and allows construction of the ROC curve, $ROC_{s,t}^{C/D}$, where $s$ is |
| | the "baseline" time of a subsetted dataset, i.e. $T \geq s$. One can specify `AT=`$t$ in the `ROCOPTIONS` in the `PROC PHREG` statement, in order to define the window over which cases accumulate, so that $T \leq t$ defines cases and $T > t$ defines controls. Specifying PLOTS=ROC in the `PROC PHREG` statement displays the ROC curve at selected time points. |
| • AUC function | Using the same options as above, but instead specifying `PLOTS=AUC` in the `PROC PHREG` statement displays the AUC and the 95% confidence limits with respect to time. |
| • Example | The SAS User's Guide for the `PHREG` procedure demonstrates the above functionality on the Mayo PBC dataset to assess time-varying performance and to compare models. |
| **Incident cases/Dynamic controls (Semiparametric estimation)** | R package `risksetROC` |
| • ROC function | `risksetROC()` calculates estimates and associated 95% confidence intervals for $ROC_t^{I/D}(p)$ by accommodating updated marker values by using time-dependent data and appropriately specifying the `entry` and `Stime` arguments. For example, consider the illustrative dataset in Table 2(a) with marker values measured only at baseline. Compare this to the time-dependent dataset in Table 2(b) that includes monthly updated marker values. When a new marker value is available, the individual is censored with the old value and re-enters the study with the new value at the updated entry time. |
| • AUC function | `risksetROC()` (described above) also calculates estimates and associated 95% confidence intervals for $AUC^{I/D}(t)$. |
| • c-index function | `risksetAUC()` estimates the c-index. Confidence intervals |
| | can be computed using bootstrapping, as illustrated in the annotated code of Appendix B. |
| • Example | The documentation for the `riskROC` package demonstrates the above functionality on a lung cancer dataset (also freely available in R, like the Mayo PBC dataset). |
| **Incident cases/Dynamic controls (Nonparametric estimation)** | R package `meanrankROC` |
| ROC function | `dynamicTP()` accommodates updated marker values by using time-dependent data as above, and appropriately specifying `start` and `stop` times for intervals with updated marker values. `dynamicTP()`, along with `nne_TPR()` provides a smooth curve over time of sensitivity (or TPF) or $ROC_t^{I/D}(p)$ for a fixed specificity 1-$p$. |
| AUC function | `MeanRank()` accommodates updated marker values by using time-dependent data as above, and appropriately specifying `start` and `stop` times for intervals with updated marker values. `MeanRank()`, along with `nne.Crossvalidate()` provides a smooth curve of $AUC^{I/D}(t)$ over time. |
| c-index function | `dynamicIntegrateAUC()` estimates the c-index. Confidence intervals can be computed using bootstrapping, as illustrated in the annotated code of Appendix B. |
| Example | See Section 4 of this tutorial (and Appendix B for corresponding R code) for an illustration of the `meanrankROC` package applied to assessing time-dependent discrimination accuracy of both baseline and time-varying markers. |

Table 2: An illustration of datasets with marker values (a) measured only at baseline and (b) updated approximately every month. Subjects are censored when a new marker value is available, and they re-enter the study with the new marker value and an updated start time.

(a) Marker measured at baseline only

| Subject | Marker | Start time (days) | Stop time (days) | Death observed |
|---------|--------|-------------------|------------------|----------------|
| 1 | $m_0$ | 0 | 65 | 1 |
| 2 | $m_0$ | 0 | 40 | 0 |

(b) Marker measured approximately monthly

| Subject | Marker | Start time (days) | Stop time (days) | Death observed |
|---------|--------|-------------------|------------------|----------------|
| 1 | $m_0$ | 0 | 25 | 0 |
| 1 | $m_{25}$ | 25 | 58 | 0 |
| 1 | $m_{58}$ | 58 | 65 | 1 |
| 2 | $m_0$ | 0 | 30 | 0 |
| 2 | $m_{30}$ | 30 | 40 | 0 |

Table 3: Time-varying performance of baseline and updated risk scores from the 4-covariate and 5-covariate models using $AUC^{I/D}$ and $AUC^{C/D}$ mimicking landmark analysis

| | $AUC^{I/D}(t)$ (95% CI) | | | c-index | $AUC^{C/D}(t, t+1\ year)$ (95% CI) | | |
|---|---|---|---|---|---|---|---|
| | t = 1 year | t = 4 years | t = 6 years | (95% CI) | t = 1 year | t = 4 years | t = 6 years |
| Baseline risk scores | | | | | | | |
| 4-covariate model | 0.84 | 0.69 | 0.64 | 0.72 | 0.77 | 0.72 | 0.77 |
| | (0.79, 0.89) | (0.60, 0.76) | (0.55, 0.70) | (0.66, 0.74) | (0.56, 0.95) | (0.55, 0.87) | (0.60, 0.88) |
| 5-covariate model | 0.88 | 0.85 | 0.66 | 0.79 | 0.80 | 0.78 | 0.65 |
| | (0.80, 0.91) | (0.74, 0.86) | (0.62, 0.78) | (0.76, 0.83) | (0.57, 0.93) | (0.66, 0.91) | (0.44, 0.89) |
| Updated risk scores | | | | | | | |
| 4-covariate model | 0.90 | 0.86 | 0.84 | 0.86 | 0.79 | 0.81 | 0.84 |
| | (0.86, 0.96) | (0.80, 0.91) | (0.77, 0.90) | (0.80, 0.89) | (0.61, 0.95) | (0.63, 0.91) | (0.63, 0.95) |
| 5-covariate model | 0.92 | 0.92 | 0.88 | 0.89 | 0.82 | 0.84 | 0.87 |
| | (0.88, 0.96) | (0.86, 0.95) | (0.82, 0.93) | (0.84, 0.92) | (0.70, 0.94) | (0.68, 0.94) | (0.66, 0.99) |

**Figure Legends**

Figure 1: An illustration of assessments at sequential baseline time points. Solid circles represent events and hollow circles represent censored subjects. At each starting time point, subjects that remain event-free are used for analysis. The solid red vertical line represents this cut-off. The dashed blue vertical line represents the subsequent 1-year cut-off which is used to define cases versus controls.

Figure 2: Time-varying prognostic accuracy of baseline risk scores obtained from the 4-covariate model versus the 5-covariate model over time using the incident/dynamic approach, with respect to (a) $AUC^{I/D}(t)$ and (b) $ROC_t^{I/D}$ for a fixed false positive fraction (FPF) of 10% (or sensitivity for a fixed specificity of 90%).

Figure 3: Time-varying prognostic accuracy of updated risk scores obtained from the 4-covariate model versus the 5-covariate model over time using the incident/dynamic approach, with respect to (a) $AUC^{I/D}(t)$ and (b) $ROC_t^{I/D}$ for a fixed false positive fraction (FPF) of 10% (or sensitivity for a fixed specificity of 90%).

Figure 4: Time-varying prognostic accuracy with 95% confidence intervals of (a) baseline and (b) updated risk scores obtained from the 5-covariate model using the incident/dynamic approach.

## Figures

### Figure 1

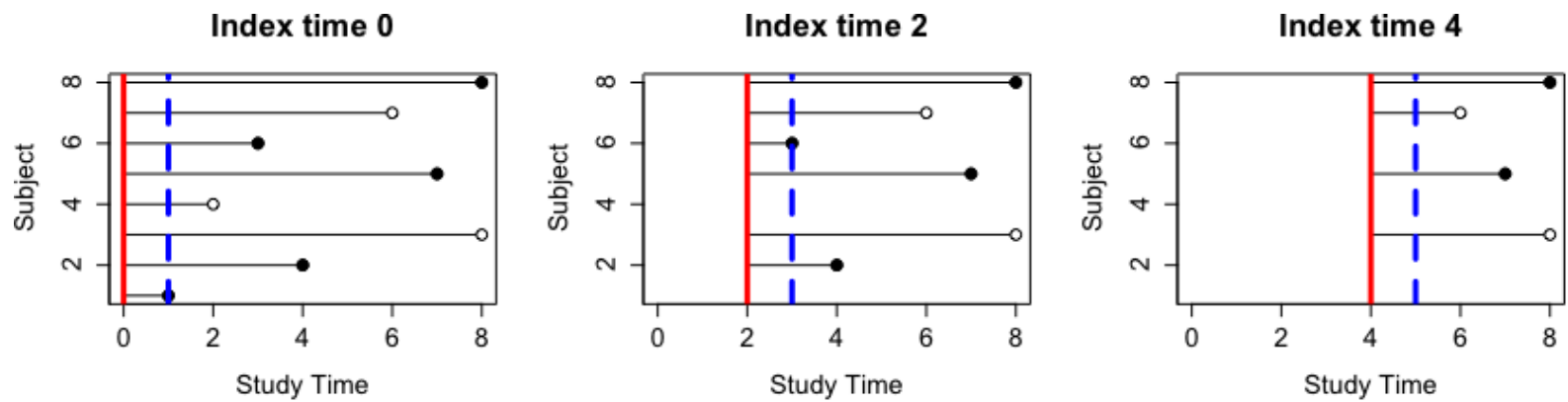

### Figure 2

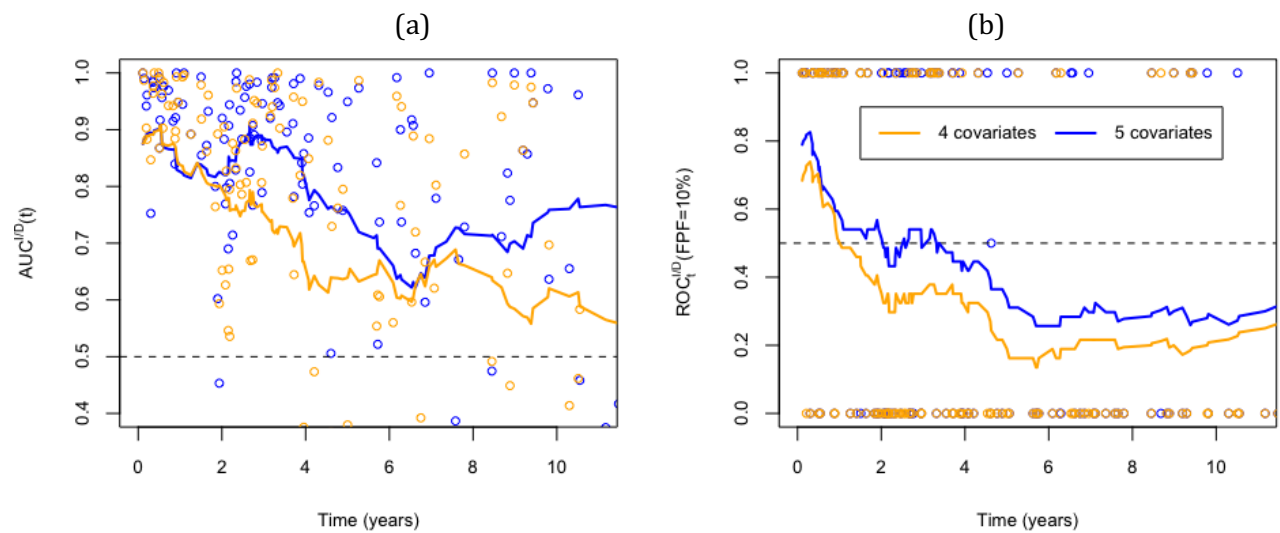

Figure 3

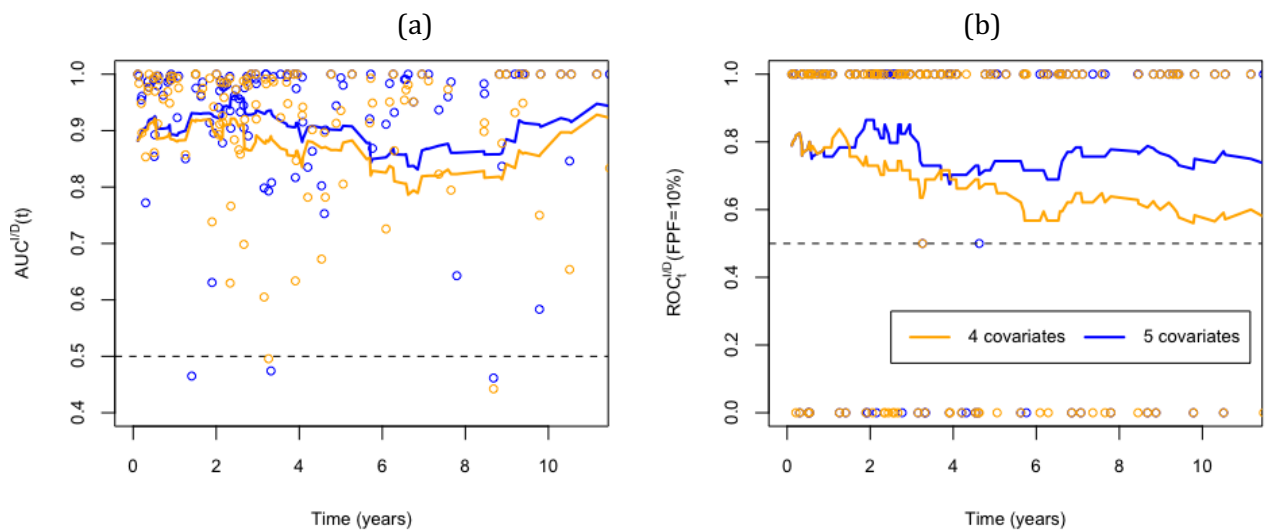

Figure 4

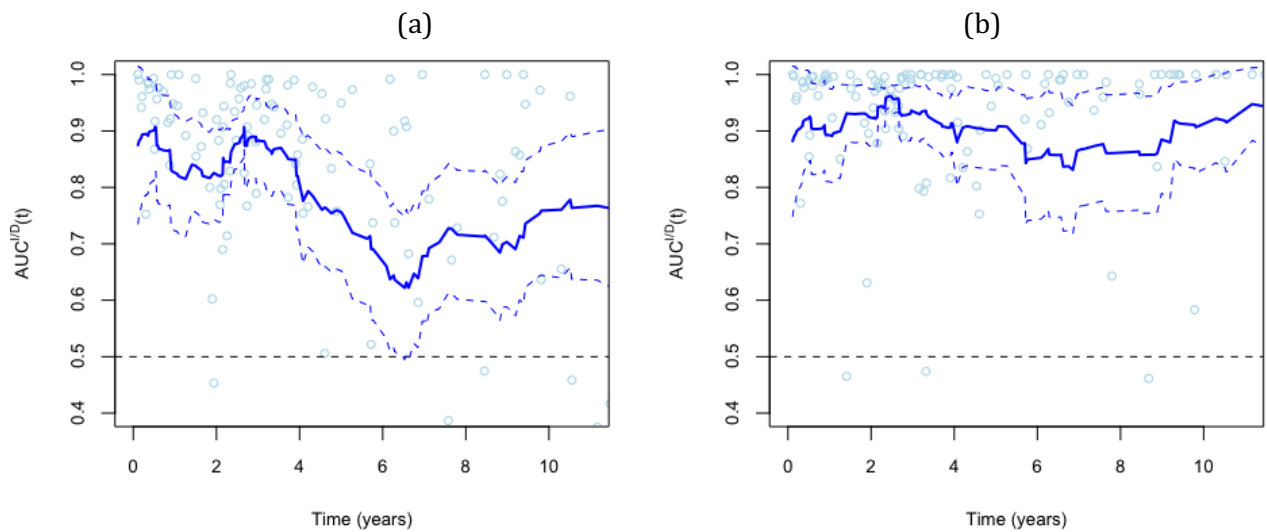

# Appendix A: Estimation Methods

Let

- $M$ denote a continuous marker or test. By convention, higher values of $M$ are more indicative of the adverse outcome
- $T$ denote the failure time
- $C$ denote the censoring time
- $Z = min(T, C)$ is the follow-up time
- $\delta$ denote the event indicator with $\delta = 1$ if $T \leq C$ and $\delta = 0$ if $T > C$
- subscript $i$ denote the variables for a subject $i$

## 1. Cumulative (Prevalent) Cases / Dynamic Controls

Let

- $s$ denote the start time of case ascertainment (often $s$=0 for baseline)
- $t$ denote the stop time of case ascertainment

At any give times $s$ and $t$ and given cut-off value $c$, we define sensitivity and specificity as:

$$\text{Se}^\text{C}(c \mid \text{start} = s, \text{stop} = t) = P(M > c \mid T \geq s , T \leq t)$$

$$\text{Sp}^\text{D}(c \mid \text{start} = s, \text{stop} = t) = P(M \leq c \mid T \geq s , T > t)$$

Using these definitions, the corresponding ROC curve can be defined at any times s, t. Heagerty et al.[1] developed two estimators for sensitivity and specificity where case ascertainment was assumed to begin at baseline, i.e. $s = 0$. These methods, described below, can be extended to sequential baseline values of $s$ to characterize time-varying performance, as described in the main text.

## (a) Kaplan-Meier estimator

Using Bayes' Theorem, the widely used nonparametric Kaplan-Meier estimate of the survival function, and the empirical distribution function of the marker $M$, Heagerty et al.[1] provided simple estimators for sensitivity and specificity as

$$\widehat{\text{Se}}^\text{C}(c \mid \text{start} = 0, \text{stop} = t) = \frac{\left\{1 - \hat{S}_{KM}(t \mid M > c)\right\}\left\{1 - \hat{F}_M(c)\right\}}{1 - \hat{S}_{KM}(t)}$$

$$\widehat{\text{Sp}}^D(c|\text{start} = 0, \text{stop} = t) = \frac{\hat{S}_{KM}(t|M \leq c)\ \hat{F}_M(c)}{\hat{S}_{KM}(t)}$$

where $\hat{S}_{KM}(t)$ is the Kaplan-Meier estimate of the survival function, $\hat{S}_{KM}(t|M > c)$ is the Kaplan-Meier estimate of the conditional survival function for the subset defined by $M > c$, and $\hat{F}_M(c) = \frac{1}{n}\sum 1(M_i \leq c)$ is the empirical distribution function of the marker $M$.

The Kaplan-Meier estimator is a standard and widely-used nonparametric estimator of the survival function, which uses all the information in the data, including censored observations, for estimation. However, there are two potential drawbacks of this estimation approach: (i) it does not guarantee that sensitivity and specificity are monotone in $M$ and bounded by [0,1], and (ii) the conditional Kaplan-Meier estimator $\hat{S}_{KM}(t|M > c)$ assumes that the censoring mechanism does not depend on $M$. This assumption may be violated in practice when the intensity of follow-up is influenced by the marker measurements, a common scenario that results in marker-dependent censoring.

## (b) Nearest neighbor estimator

An alternative approach proposed by Heagerty et al.[1] to address the above drawbacks is based on a nearest neighbor estimator for the bivariate distribution function of $(M, T)$, $F(c,t) = P(M \leq c, T \leq t)$, or equivalently, $S(c,t) = P(M > c, T > t)$, that was provided by Akritas[2]. The estimator is based on the representation: $S(c,t) = \int_c^\infty S(t|M = s)dF_M(s)$, where $F_M(s)$ is the distribution function of $M$. This estimator is provided by

$$\hat{S}_{\lambda_n}(c, t) = \frac{1}{n}\sum_i \hat{S}_{\lambda_n}(t|M = M_i)1(M_i > c),$$

where $\hat{S}_{\lambda_n}(t|M = M_i)$ is a suitable estimator of the conditional survival function characterized by a smoothing parameter $\lambda_n$. Unless $M$ is discrete and there are sufficient observations at each value of $M$, some smoothing is required to estimate $S(t | M = M_i)$. $K_{\lambda_n}(M_j, M_i)$ is defined as a kernel function that depends on a smoothing parameter $\lambda_n$. Using the kernel function, a weighted Kaplan-Meier estimator follows:

$$\hat{S}_{\lambda_n}(t|M = M_i) = \prod_{s \in \mathcal{T}_n, s \leq t}\left\{1 - \frac{\sum_j K_{\lambda_n}(M_j, M_i)\,1(Z_j = s)\delta_j}{\sum_j K_{\lambda_n}(M_j, M_i)\,1(Z_j = s)}\right\}$$

where $\mathcal{T}_n$ is the set of unique values of $Z_i$ for observed events, $\delta_i = 1$.

Akritas[2] used a 0/1 nearest neighbor kernel, $K_{\lambda_n}(M_j, M_i) = 1\{-\lambda_n < \hat{F}_M(M_i) - \hat{F}_M(M_j) < \lambda_n\}$, where $2\lambda_n \in (0, 1)$ represents the percentage of individuals that are included in each neighborhood. The resulting estimates of sensitivity and specificity are given by

$$\widehat{Se}^C(c|start = 0, stop = t) = \frac{\{1 - \hat{F}_X(c)\} - \hat{S}_{\lambda_n}(c, t)}{1 - \hat{S}_{\lambda_n}(t)}$$

$$\widehat{Sp}^D(c|start = 0, stop = t) = 1 - \frac{\hat{S}_{\lambda_n}(c, t)}{\hat{S}_{\lambda_n}(t)}$$

where $\hat{S}_{\lambda_n}(t) = \hat{S}_{\lambda_n}(-\infty, t)$. These estimates allow for monotonicity of the sensitivity and specificity. Furthermore, since only local Kaplan-Meier estimators are used in each possible neighborhood of M=$m$, the censoring process is allowed to depend on the marker $M$.

## 2. Incident Cases / Dynamic Controls

At any give time $t$ and given cut-off value $c$, incident sensitivity and dynamic specificity are defined by dichotomizing the risk set at time $t$ into those observed to die (cases) and those observed to survive (controls):

$$Se^I(c \mid t) = P(M > c \mid T = t)$$
$$Sp^D(c \mid t) = P(M \le c \mid T > t)$$

Using these definitions, the corresponding ROC curve can be defined at any time $t$. Below we describe two estimators for sensitivity and specificity based on the incident/dynamic definition.

## (a) Semi-parametric Cox model based estimator

Heagerty & Zheng[3] proposed Cox model based methods that use riskset reweighting based on the estimated hazard in order to estimate sensitivity and specificity. The censoring time is assumed to be independent of the failure time and marker. Under proportional hazards, a standard Cox model is fit:

$$\lambda(t \mid M_i) = \lambda_0(t) \, exp(M_i \, \gamma)$$

To estimate sensitivity and specificity, i.e. the marker distribution conditional on survival time, Heagerty & Zheng[3] use Xu and O'Quigley's result that partial likelihood estimation methods can be exploited to provide model-based estimates of the distribution of covariates conditional on survival time.[3,4] Specifically, letting $R_i(t) = 1(M_i \geq t)$ denote the at-risk indicator, $\pi_i(\gamma, t) = \frac{R_i(t)\,\exp(M_i\gamma)}{\sum_j R_j(t)\,\exp(M_j\gamma)}$ can be used to estimate the distribution of marker $M$, conditional on the event occurring at time $t$, so that $\hat{P}(M_i \leq m \mid T_i = t) = \sum_k \pi_k(\hat{\gamma}, t)\ 1(M_k \leq m)$. This result and using partial likelihood to estimate $\gamma$ directly give a semiparametric estimator of sensitivity, which uses a reweighting of the marker distribution observed among the riskset at a time $t$:

$$\widehat{Se}^{\mathrm{I}}(c \mid t) = \hat{P}(M > c \mid T = t) = \sum_k 1(M_k > c)\ \pi_k(\hat{\gamma}, t)$$

The methods can also accommodate non-proportional hazards. A varying-coefficient model of the form $\lambda(t \mid M_i) = \lambda_0(t)\ exp(M_i\,\gamma(t))$ can be fit to obtain the time-varying coefficient $\hat{\gamma}(t)$ and estimate sensitivity as:

$$\widehat{Se}^{\mathrm{I}}(c \mid t) = \hat{P}(M_i > c \mid T_i = t) = \sum_k 1(M_k > c)\ \pi_k[\hat{\gamma}(t), t]$$

The time-varying coefficient, $\gamma(t)$, and subsequently $AUC^{\mathrm{I/D}}(t)$, can be estimated using flexible semiparametric locally weighted partial likelihood methods[5] or local linear smoothing of the scaled Schoenfeld residuals.

An empirical estimator of specificity is given as:

$$\widehat{Sp}^{D}(c \mid t) = \hat{P}(M \leq c \mid T > t) = \frac{\sum_k 1(M_k > c, T_k > t)}{\sum_k 1(T_k > t)}$$

## (b) Non-parametric rank-based estimator

A nonparametric rank-based approach for the estimation of $AUC^{\mathrm{I/D}}(t)$ was proposed by Saha-Chaudhuri & Heagerty[6]. For a fixed time $t$, a percentile is calculated for each case in the risk set relative to the controls in the risk set. A perfect marker would have the case marker value greater than 100% of risk set controls. The mean percentile at time $t$ is calculated as the mean of the percentiles for all cases at $t$, as follows:

$$A(t) = \frac{1}{n_t d_t} \sum_{i \in \mathcal{R}_t^1} \sum_{j \in \mathcal{R}_t^0} 1(M_i > M_j)$$

where $\mathcal{R}_t^1$ and $\mathcal{R}_t^0$ denote the sets of cases and controls in the risk set at time $t$, respectively.

Unless there are sufficient events at each time $t$, some smoothing is typically required to estimate the AUC. Using a standardized kernel function such that $\sum_j K_{h_n}(t - t_j) = 1$ based on a neighborhood of points defined by parameter $h_n$, Saha-Chaudhuri & Heagerty[6] defined a smoothed estimator of AUC by:

$$\widehat{AUC}(t) = \sum_j K_{h_n}(t - t_j) \, A(t_j).$$

They used a nearest neighbor kernel, resulting in the following weighted mean rank (WMR) estimator:

$$WMR(t) = \frac{1}{|\mathcal{N}_t(h_n)|} \sum_{t_j \in \mathcal{N}_t(h_n)} A(t_j)$$

where $\mathcal{N}_t(h_n) = (t_j : |t - t_j| < h_n)$ denotes a neighborhood around time $t$. This estimator is used to estimate the summary curve, $\text{AUC}^{\text{I/D}}(t)$, as the local average of mean case percentiles. This nonparametric approach provides a simple description of marker performance within each risk set and, by smoothing individual case percentiles, a final summary curve characterizes accuracy over time.

A smooth curve of sensitivity for a fixed specificity can be estimated in a similar manner as:

$$\widehat{Se}^{\text{I}}(\text{Sp} \mid t) = \sum_j K_{h_n}(t - t_j) \, 1[A(t) > \text{Sp}]$$

**(c) The concordance-index or c-index**

The c-index can also be expressed as a weighted average of the area under time-specific ROC curves (AUCs)[3], obtained using the incident/dynamic definition of sensitivity and specificity:

$$\text{c-index} = \int_t \text{AUC}^{\text{I/D}}(t) \, \text{w}(t) \, dt$$

where w(t)=2 f(t) S(t), f(t) represents the distribution of failure times $T$ and S(t) represents the survival time. The c-index is straightforward to estimate using the methods described above. Specifically, $\text{AUC}^{\text{I/D}}(t)$ can be estimated using the semiparametric or nonparametric approaches described in subsections (a) and (b) above, respectively, and f(t) and S(t) are derived nonparametrically, using the Kaplan-Meier estimate of the survival function.

## 3. Competing Risk Outcomes

Here we assume that a single event time $T_i$ may correspond to $J$ mutually exclusive types or causes of failure, $j = 1, 2, ..., J$ and we may be interested in one or more specific types. We generalize the definition of the event indicator $\delta_i$, so that $\delta_i = 1, 2, ..., J$ indicates a specific type or cause of failure, while $\delta_i = 0$ indicates censoring as before.

### a) Cumulative (Prevalent) Cases / Dynamic Controls

For the setting of competing risk events, Saha & Heagerty[7] modified the approach of Heagerty et al[1] by using nearest neighbor estimation for the cumulative incidence function (CIF) associated with each type of failure, instead of the bivariate distribution function of the marker and time, $(M, T)$. Estimation of sensitivity is based on the weighted conditional CIF, estimated as follows:

$$\hat{C}_j(t|M = M_i) = \sum_{s < t} \hat{S}_{\in_n}(s|M = M_i)\hat{\lambda}_j(s|M = M_i),$$

where $\hat{\lambda}_j(s|M = M_i)$ is the observed hazard for event type $j$ at time $t$ and $\hat{S}_{\in_n}(s|M = M_i)$ is a locally weighted Kaplan-Meier estimator of the conditional survival function, defined as before using a nearest neighbor kernel $K_{\in_n}(M_j, M_i)$ that depends on a smoothing parameter $\in_n$, with $2\in_n \in (0, 1)$ representing the percentage of individuals that are included in each neighborhood. Using the kernel function, a weighted Kaplan-Meier estimator follows:

$$\hat{S}_{\in_n}(t|M = M_i) = \prod_{s \in \mathcal{T}_n, s \leq t} \left\{ 1 - \frac{\sum_k K_{\in_n}(M_k, M_i)\, 1(Z_k = s)\delta_k}{\sum_k K_{\in_n}(M_k, M_i)\, 1(Z_k \geq s)} \right\}$$

where $\mathcal{T}_n$ is the set of unique observed event times for the event of interest, $\delta = j$. The resulting estimate of sensitivity for event $j$ is given by

$$\widehat{Se}^C_j(c|\text{start} = 0, \text{stop} = t) = \frac{P(M > c, T \leq t, event\ type = j)}{P(T \leq t, event\ type = j)} = \frac{\int_c^\infty \hat{C}_j(t|M = u)\hat{f}_M(u)du}{\int_{-\infty}^\infty \hat{C}_j(t|M = u)\hat{f}_M(u)du}$$

where $f_M(s)$ is the probability density function of marker $M$.

To estimate specificity, Saha & Heagerty[7] also use the CIF conditional on marker $M$ to get:

$$\widehat{Sp}^D(c|\text{start} = 0, \text{stop} = t) = \frac{\hat{P}(M > c, T > t)}{\hat{P}(T > t)} = \frac{\int_c^\infty \hat{P}(T > t, M = u)du}{\int_{-\infty}^\infty \hat{P}(T > t, M = u)du}$$

$$= \frac{\int_c^\infty \hat{P}(T > t | M = u) \hat{f}_M(u) du}{\int_{-\infty}^\infty \hat{P}(T > t | M = u) \hat{f}_M(u) du}$$

$$= \frac{\int_c^\infty \left[1 - \sum_j \hat{P}(T \leq t, \delta = j \mid M = m)\right] \hat{f}_M(u) du}{\int_{-\infty}^\infty \left[1 - \sum_j \hat{P}(T \leq t, \delta = j \mid M = m)\right] \hat{f}_M(u) du}$$

$$= \frac{\int_c^\infty \left[1 - \sum_j \hat{C}_j(t \mid M = m)\right] \hat{f}_M(u) du}{\int_{-\infty}^\infty \left[1 - \sum_j \hat{C}_j(t \mid M = m)\right] \hat{f}_M(u) du}$$

## b) Incident Cases / Dynamic Controls

Saha & Heagerty[7] showed that the riskset reweighting used by Heagerty & Zheng[3] to estimate the sensitivity $P(M_i > c \mid T_i = t)$ can also be used with competing risks data. Under proportional hazards, a standard Cox model is fit for event of type $j$:

$$\lambda_j(t \mid M_i) = \lambda_{0,j}(t) \, exp(M_i \gamma_j)$$

where $\gamma_j$ is the cause-specific hazard for event of type $j$ associated with the marker. $\gamma_j$ can be estimated using Maximum Partial Likelihood Estimation by censoring all other types of failure. As before, letting $R_i(t) = 1(M_i \geq t)$ denote the at-risk indicator for time $t$, $\pi_i^j(\gamma_j, t) = \frac{R_i(t) \exp(M_i \gamma_j)}{\sum_k R_k(t) \exp(M_k \gamma_j)}$ for the event of type $j$ can be used to estimate the distribution of marker $M$, conditional on event $j$ occurring at time $t$, so that the estimates of sensitivity and specificity are analogous to those presented by Heagerty & Zheng[3]. Specifically, we get the following semiparametric estimator of sensitivity for event type $j$:

$$\widehat{Se}^I_j(c \mid t) = \hat{P}(M > c \mid T = t, \text{event type} = j) = \sum_k 1(M_k > c) \, \pi_k^j(\hat{\gamma}_j, t).$$

The methods can also accommodate non-proportional hazards, by replacing $\hat{\gamma}_j$ with an estimate of the time-varying coefficient, $\gamma_j(t)$, just as before. The time-varying coefficient, $\gamma_j(t)$, and subsequently $AUC^{I/D}(t)$, can be estimated using flexible semiparametric locally weighted partial likelihood methods[5] or local linear smoothing of the scaled Schoenfeld residuals.

An empirical estimator of specificity is given as

$$\widehat{Sp}^D(c \mid t) = \hat{P}(M \leq c \mid T > t) = \frac{\sum_k 1(M_k > c, T_k > t)}{\sum_k 1(T_k > t)}$$

## Appendix B: Annotated R code

```r
library(survival)

install.packages("survivalROC")
install.packages("risksetROC")
library(survivalROC)
library(risksetROC)

#Download meanrankROC package from github: https://github.com/aasthaa/meanrankROC_package

source("MeanRank.q")
source("NNE-estimate.q")
source("NNE-CrossValidation.q")
source("interpolate.q")

source("dynamicTP.q")
source("NNE-estimate_TPR.q")

source("dynamicIntegrateAUC.R")

#Load in the datasets. Note: The PBC data is freely available in R.
bDat <- pbc[1:312,] #baseline data
bDat$deathEver <- bDat$status
bDat$deathEver[which(bDat$status==1)] <- 0 #censor at transplant
bDat$deathEver[which(bDat$status==2)] <- 1 #death

#Build dataset with time-dependent covariates
pbc2 <- tmerge(pbc, pbc, id=id, death = event(time, status)) #set range
pbc2 <- tmerge(pbc2, pbcseq, id=id, ascites = tdc(day, ascites), hepato = tdc(day, hepato),
    spiders = tdc(day, spiders), edema = tdc(day, edema), chol = tdc(day, chol),
    bili = tdc(day, bili), albumin = tdc(day, albumin),
    protime = tdc(day, protime), alk.phos = tdc(day, alk.phos),
    ast = tdc(day, ast), platelet = tdc(day, platelet), stage = tdc(day, stage) )
```

```r
length(unique(pbc2$id))

dim(pbc2)
pbc2 <- subset(pbc2, id>=1 & id<=312)
dim(pbc2)
length(unique(pbc2$id))
dim(bDat)

#According to documentation, some baseline values for protime and age in pbc were found to be incorrect.
 Correct values in pbcseq
bDat[1:5,]
subset(pbc2, tstart==0)[1:5,]
for(i in 1:312){
    if(pbc2$protime[which(pbc2$id==i & pbc2$tstart==0)] != bDat$protime[i])
        pbc2$protime[which(pbc2$id==i & pbc2$tstart==0)] <- bDat$protime[i]

    if(pbc2$age[which(pbc2$id==i & pbc2$tstart==0)] != bDat$age[i])
        pbc2$age[which(pbc2$id==i & pbc2$tstart==0)] <- bDat$age[i]
}

pbc2$deathEver <- pbc2$status
pbc2$deathEver[which(pbc2$status==1)] <- 0
pbc2$deathEver[which(pbc2$status==2)] <- 1

pbc2$death[which(pbc2$death==1)] <- 0
pbc2$death[which(pbc2$death==2)] <- 1

#Use 10-fold CV to get baseline scores
set.seed(49)

samples <- floor(runif(nrow(bDat), 1,11))
sampSizes <- sapply(seq(1:10), function(s){length(which(samples==s))} )
```

```
sampSizes  #Check that no subsets with 0 subjects

while(min(sampSizes)==0) {
    samples <- floor(runif(nTrain, 1,11))
    sampSizes <- sapply(seq(1:10), function(s){length(which(samples==s))} )
}

###10-fold cross-validation to get predicted baseline scores
score4Baseline_cv <- score5Baseline_cv <- rep(NA,nrow(bDat))

for(s in 1:10) {
    bDat_train <- bDat[-which(samples==s),]
    bDat_test <- bDat[which(samples==s),]

    mod <- coxph(Surv(time=time, event= deathEver) ~ log(bili) + log(protime) + edema + albumin + age,
        data=bDat_train )
    riskVals <- predict(mod, type="risk", newdata= bDat_test)
    score5Baseline_cv[which(samples==s)] <- riskVals

    mod <- coxph(Surv(time=time, event= deathEver) ~ log(protime) + edema + albumin + age, data=bDat_train )
    riskVals <- predict(mod, type="risk", newdata= bDat_test)
    score4Baseline_cv[which(samples==s)] <- riskVals
}
bDat$score4baseline <- score4Baseline_cv
bDat$score5baseline <- score5Baseline_cv

#Fit model on all baseline data, use for prediction of time-varying score
coxMod5baseline <- coxph(Surv(time=time, event= deathEver) ~ log(bili) + log(protime) + edema + albumin +
 age, data= bDat)
pbc2$score5tv <- predict(coxMod5baseline, type="risk", newdata= pbc2)

coxMod4baseline <- coxph(Surv(time=time, event= deathEver) ~ log(protime) + edema + albumin + age, data=
 bDat)
pbc2$score4tv <- predict(coxMod4baseline, type="risk", newdata= pbc2)
```

```
#####Table 2
##A. AUC_I/D
tableAUC_ID <- matrix(nrow=2, ncol=length(landmarkTimes))
tableAUC_TV_ID <- matrix(nrow=2, ncol=length(landmarkTimes))

#Baseline risk scores
scores <- c("score4baseline","score5baseline")
for(i in 1:length(scores)) {
    currVar <- eval(parse(text=paste("bDat$", scores[i],sep="")))
    mmm <- MeanRank( survival.time= bDat$time, survival.status= bDat$deathEver, marker= currVar )
    bandwidths <- 0.05 + c(1:80)/200
    IMSEs <- vector(length=length(bandwidths))
    for(j in 1:length(bandwidths)) {
        nnnC <- nne.CrossValidate(x=mmm$time, y=mmm$mean.rank, lambda=bandwidths[j]) #CV bandwidth
        IMSEs[j] <- nnnC$IMSE
    }
    currLambdaOS <- mean(bandwidths[which(IMSEs==min(IMSEs, na.rm=T))])
    nnn <- nne(x= mmm$time, y= mmm$mean.rank, lambda=currLambdaOS, nControls=mmm$nControls) #Fixed bandwidth
    tableAUC_ID[i,] <- sapply(landmarkTimes, function(x){ interpolate( x = nnn$x, y=nnn$nne, target=x ) } )
}
rownames(tableAUC_ID) <- scores
colnames(tableAUC_ID) <- landmarkTimes/units
round(tableAUC_ID, 2)

#Updated (time-varying) risk scores
scores <- c("score4tv", "score5tv")
for(i in 1:length(scores)) {
    currVar <- eval(parse(text=paste("pbc2$", scores[i],sep="")))
    mmm <- MeanRank(survival.time=pbc2$tstop, survival.status=pbc2$death, marker=currVar, start=pbc2$tstart)
    bandwidths <- 0.05 + c(1:80)/200
    IMSEs <- vector(length=length(bandwidths))
    for(j in 1:length(bandwidths)) {
        nnnC <- nne.CrossValidate(x=mmm$time, y=mmm$mean.rank, lambda=bandwidths[j]) #CV bandwidth
```

```
        IMSEs[j] <- nnnC$IMSE
    }
    currLambdaOS <- mean(bandwidths[which(IMSEs==min(IMSEs, na.rm=T))])
    nnn <- nne(x=mmm$time, y=mmm$mean.rank, lambda=currLambdaOS, nControls=mmm$nControls )   #Fixed bandwidth
    tableAUC_TV_ID[i,] <- sapply(landmarkTimes, function(x){ interpolate( x = nnn$x, y=nnn$nne, target=x ) }
 )
}
rownames(tableAUC_TV_ID) <- scores
colnames(tableAUC_TV_ID) <- landmarkTimes/units
round(tableAUC_TV_ID, 2)

#B. c-index
round(dynamicIntegrateAUC(survival.time=bDat$time, survival.status=bDat$deathEver,
 marker=bDat$score4baseline, cutoffTime = units*10), 2)
round(dynamicIntegrateAUC(survival.time=bDat$time, survival.status=bDat$deathEver,
 marker=bDat$score5baseline, cutoffTime = units*10), 2)

round(dynamicIntegrateAUC(survival.time= pbc2$tstop, survival.status= pbc2$death, start=pbc2$tstart,
 marker=pbc2$score4tv, cutoffTime = units*10), 2)
round(dynamicIntegrateAUC(survival.time= pbc2$tstop, survival.status= pbc2$death, start=pbc2$tstart,
 marker=pbc2$score5tv, cutoffTime = units*10), 2)

#C. Sequential C/D AUCs on subsetted data at each timepoint and one year ahead to mimic landmark analysis
units <- 365.25

landmarkTimes <- c(1, 4, 6)*units
tableAUC_CD <- matrix(nrow=4, ncol=length(landmarkTimes))

timeWindow <- 1

for(j in 1:length(landmarkTimes)) {
    currData <- subset(bDat, time >= (landmarkTimes[j]))
    currDataTV <- subset(pbc2, tstart <= (landmarkTimes[j]) & tstop > (landmarkTimes[j]))

    nobs <- nrow(currData)
```

```r
    out1 <- survivalROC( currData$time, currData$deathEver, marker= currData$score4baseline,
            predict.time=(landmarkTimes[j] + timeWindow*units), method="NNE", span=0.04*nobs^(-0.2))
    tableAUC_CD[1,j] <- out1$AUC

    out1 <- survivalROC( currData$time, currData$deathEver, marker= currData$score5baseline,
            predict.time=(landmarkTimes[j] + timeWindow*units), method="NNE", span=0.04*nobs^(-0.2))
    tableAUC_CD[2,j] <- out1$AUC

    nobs <- nrow(currDataTV)
    out1 <- survivalROC( currDataTV$time, currDataTV$deathEver, marker= currDataTV$score4tv,
            predict.time=(landmarkTimes[j] + timeWindow*units), method="NNE", span=0.04*nobs^(-0.2))
    tableAUC_CD[3,j] <- out1$AUC

    out1 <- survivalROC( currDataTV$time, currDataTV$deathEver, marker= currDataTV$score5tv,
            predict.time=(landmarkTimes[j] + timeWindow*units), method="NNE", span=0.04*nobs^(-0.2))
    tableAUC_CD[4,j] <- out1$AUC
}
rownames(tableAUC_CD) <- c("score4baseline","score5baseline", "score4tv", "score5tv")
colnames(tableAUC_CD) <- landmarkTimes/units
round(tableAUC_CD, 2)

####Bootstrap 95% CIs
nBoot <- 500

##A. Bootstrap CIs - Baseline markers/scores
markers <- c("score4baseline","score5baseline")
set.seed(49)
Cindex_bstrap <- matrix(nrow=nBoot, ncol=length(markers))
bstrapRes <- list()

for(b in 1:nBoot) {
    currData <- bDat[sample(x=seq(1,nrow(bDat)), size=nrow(bDat), replace = T),]
    kmfit <- survfit(Surv(time, deathEver) ~ 1, data= currData)
```

```r
currDataLM1 <- currData[which(currData$time >= (landmarkTimes[1])), ]
currDataLM2 <- currData[which(currData$time >= (landmarkTimes[2])), ]
currDataLM3 <- currData[which(currData$time >= (landmarkTimes[3])), ]

aucID_scores <- NULL
aucCD_scores <- matrix(nrow=length(scores), ncol=length(landmarkTimes))

for(i in 1:length(markers)) {
    currVar <- eval(parse(text=paste("currData$",markers[i],sep="")))

    ### AUC I/D
    mmm <- MeanRank( survival.time= currData$time, survival.status= currData$deathEver, marker= currVar )
    nnn <- nne( x= mmm$time, y= mmm$mean.rank, lambda=0.3 )  #Fixed bandwidth
    aucID_scores <- rbind(aucID_scores,
        sapply(landmarkTimes, function(t){ interpolate( x = nnn$x, y=nnn$nne, target=t ) } ))

    ### C-index
    Cindex_bstrap[b,i] <- dynamicIntegrateAUC(survival.time=currData$time,
        survival.status= currData$deathEver, marker=currVar, cutoffTime = units*10)

    ### AUC C/D landmark
    if(markers[i]=="score4baseline" | markers[i]=="score5baseline") {
        currDataLM <- currDataLM1
        currVecLM <- eval(parse(text=paste("currDataLM$", markers[i], sep="")))
        out1 <- survivalROC( currDataLM$time, currDataLM$deathEver, marker=currVecLM,
                predict.time=(landmarkTimes[1] + timeWindow*units), method="NNE", span=0.04*nobs^(-0.2))
        currDataLM <- currDataLM2
        currVecLM <- eval(parse(text=paste("currDataLM$", markers[i], sep="")))
        out2 <- survivalROC( currDataLM$time, currDataLM$deathEver, marker=currVecLM,
                predict.time=(landmarkTimes[2] + timeWindow*units), method="NNE", span=0.04*nobs^(-0.2))

        currDataLM <- currDataLM3
        currVecLM <- eval(parse(text=paste("currDataLM$", markers[i], sep="")))
        out3 <- survivalROC( currDataLM$time, currDataLM$deathEver, marker=currVecLM,
                predict.time=(landmarkTimes[3] + timeWindow*units), method="NNE", span=0.04*nobs^(-0.2))
```

```r
        aucCD_scores[i,] <- c(out1$AUC, out2$AUC, out3$AUC)
      }
    }
    bstrapRes[[b]] <- list(aucID_scores=aucID_scores, aucCD_scores=aucCD_scores)
}

#Get CIs for c-indices
Cindex_CIs <- round(apply(Cindex_bstrap, 2, quantile, probs=c(0.025,0.975)),2)
colnames(Cindex_CIs) <- markers
Cindex_CIs

#Get CIs for AUCs
AUC_ID_CIs <- NULL
AUC_CD_CIs <- NULL

for(t in 1:length(landmarkTimes)) {
   AUC_ID <- NULL
   AUC_CD <- NULL
   for(b in 1:nBoot) {
       AUC_ID <- cbind( AUC_ID, bstrapRes[[b]]$aucID_scores[,t] )
       AUC_CD <- cbind( AUC_CD, bstrapRes[[b]]$aucCD_scores[,t] )
   }
   AUC_ID_CI_raw <- round( apply(AUC_ID, 1, quantile, probs=c(0.025,0.975)), 2 )
   AUC_CD_CI_raw <- round( apply(AUC_CD, 1, quantile, probs=c(0.025,0.975)), 2 )

   AUC_ID_CIs <- cbind(AUC_ID_CIs, sapply(seq(2), function(x) paste("(", AUC_ID_CI_raw[1,x], ", ",
 AUC_ID_CI_raw[2,x], ")", sep="" ) ) )
   AUC_CD_CIs <- cbind(AUC_CD_CIs, sapply(seq(2), function(x) paste("(", AUC_CD_CI_raw[1,x], ", ",
 AUC_CD_CI_raw[2,x], ")", sep="" ) ) )
}
rownames(AUC_CD_CIs) <- rownames(AUC_ID_CIs) <- c("4-cov model", "5-cov model")
colnames(AUC_CD_CIs) <- colnames(AUC_ID_CIs) <- landmarkTimes/units
AUC_ID_CIs
AUC_CD_CIs
```

```
##B. Bootstrap CIs - Time-varying scores
markers <- c("score4tv","score5tv")

set.seed(49)
Cindex_bstrapTV <- matrix(nrow=nBoot, ncol=length(markers))
bstrapResTV <- list()

for(b in 1:nBoot) {
    #sample individuals
    subjs <- unique(pbc2$id)
    currSubjs <- sample(x=subjs, size=length(subjs), replace = T)
    currData <- NULL
    for(j in 1:length(currSubjs))
        currData <- rbind(currData, pbc2[which(pbc2$id==currSubjs[j]),])

    kmfit <- survfit(Surv(time=tstart, time2=tstop, event=death) ~ 1, data=currData)

    currDataLM1 <- currData[which(currData$tstart<=(landmarkTimes[1]) & currData$tstop>(landmarkTimes[1])),]
    currDataLM2 <- currData[which(currData$tstart<=(landmarkTimes[2]) & currData$tstop>(landmarkTimes[2])),]
    currDataLM3 <- currData[which(currData$tstart<=(landmarkTimes[3]) & currData$tstop>(landmarkTimes[3])),]

    aucID_scores <- NULL
    aucCD_scores <- matrix(nrow=length(scores), ncol=length(landmarkTimes))

    for(i in 1:length(markers)) {
        currVar <- eval(parse(text=paste("currData$", markers[i],sep="")))

        ### AUC I/D
        mmm <- MeanRank(survival.time=currData$tstop, survival.status=currData$death, start=currData$tstart,
            marker=currVar)
        nnn <- nne( x= mmm$time, y= mmm$mean.rank, lambda=0.3, nControls=mmm$nControls ) #Fixed bandwidth
        aucID_scores <- rbind(aucID_scores,
            sapply(landmarkTimes, function(t){ interpolate( x = nnn$x, y=nnn$nne, target=t ) } ))

        ### C-index
```

```
        Cindex_bstrapTV[b,i] <- dynamicIntegrateAUC(survival.time=currData$tstop,
            survival.status=currData$death, start=currData$tstart, marker= currVar, cutoffTime = units*10)

        ### AUC C/D landmark (for the scores only)
            currDataLM <- currDataLM1
            currVecLM <- eval(parse(text=paste("currDataLM$", markers[i], sep="")))
            out1 <- survivalROC( currDataLM$time, currDataLM$deathEver, marker=currVecLM,
                    predict.time=(landmarkTimes[1] + timeWindow*units), method="NNE", span=0.04*nobs^(-0.2))
            currDataLM <- currDataLM2
            currVecLM <- eval(parse(text=paste("currDataLM$", markers[i], sep="")))
            out2 <- survivalROC( currDataLM$time, currDataLM$deathEver, marker=currVecLM,
                    predict.time=(landmarkTimes[2] + timeWindow*units), method="NNE", span=0.04*nobs^(-0.2))
            currDataLM <- currDataLM3
            currVecLM <- eval(parse(text=paste("currDataLM$", markers[i], sep="")))
            out3 <- survivalROC( currDataLM$time, currDataLM$deathEver, marker=currVecLM,
                    predict.time=(landmarkTimes[3] + timeWindow*units), method="NNE", span=0.04*nobs^(-0.2))
            aucCD_scores[i,] <- c(out1$AUC, out2$AUC, out3$AUC)

    }
    bstrapResTV[[b]] <- list(aucID_scores, aucCD_scores)
}

#Get CIs for c-indices
Cindex_CIs <- round(apply(Cindex_bstrapTV, 2, quantile, probs=c(0.025,0.975)), 2)
colnames(Cindex_CIs) <- markers
Cindex_CIs

#Get CIs for AUCs
AUC_CD_CIs <- NULL
AUC_ID_CIs <- NULL

for(t in 1:length(landmarkTimes)) {
    AUC_CD <- NULL
    AUC_ID <- NULL
    for(b in 1:nBoot) {
```

```
        AUC_ID <- cbind( AUC_ID, bstrapResTV[[b]][[1]][,t] )
        AUC_CD <- cbind( AUC_CD, bstrapResTV[[b]][[2]][,t] )
    }
    AUC_CD_CI_raw <- round( apply(AUC_CD, 1, quantile, probs=c(0.025,0.975)), 2 )
    AUC_ID_CI_raw <- round( apply(AUC_ID, 1, quantile, probs=c(0.025,0.975)), 2 )

    AUC_CD_CIs <- cbind(AUC_CD_CIs, sapply(seq(2), function(x) paste("(", AUC_CD_CI_raw[1,x], ", ",
        AUC_CD_CI_raw[2,x], ")", sep="" ) ) )
    AUC_ID_CIs <- cbind(AUC_ID_CIs, sapply(seq(2), function(x) paste("(", AUC_ID_CI_raw[1,x], ", ",
        AUC_ID_CI_raw[2,x], ")", sep="" ) ) )
}
rownames(AUC_CD_CIs) <- rownames(AUC_ID_CIs) <- c("4-cov model", "5-cov model")
colnames(AUC_CD_CIs) <- colnames(AUC_ID_CIs) <- landmarkTimes/units
AUC_ID_CIs
AUC_CD_CIs

#C-index difference
getCindexBstrapCI <- function(nBoot, inData, markerVarName1, markerVarName2, timeVarName,
  eventVarName, cutoffTime) {
    set.seed(49)
    resultStar <- vector(length=nBoot)
    for(i in 1:nBoot) {
        datStar <- inData[sample(seq(nrow(inData)), nrow(inData), replace=T), ]

        markerVar1 <- eval(parse(text=paste("datStar$", markerVarName1, sep="")))
        markerVar2 <- eval(parse(text=paste("datStar$", markerVarName2, sep="")))
        timeVar <- eval(parse(text=paste("datStar$", timeVarName, sep="")))
        eventVar <- eval(parse(text=paste("datStar$", eventVarName, sep="")))

        kmfit <- survfit(Surv(timeVar, eventVar) ~ 1)

        ### Marker 1
        mmm <- MeanRank( survival.time= timeVar, survival.status= eventVar, marker= markerVar1 )

        #Get overlap between survival function and mmm
```

```r
    meanRanks <-  mmm$mean.rank[which(mmm$time <= cutoffTime)]
    survTimes <- mmm$time[mmm$time <= cutoffTime]
    timeMatch <- match(survTimes, kmfit$time)
    S_t <- kmfit$surv[timeMatch]

    #Calculate weights for c-index
    f_t <- c( 0, (S_t[-length(S_t)] - S_t[-1]) )
    S_tao <- S_t[length(S_t)]
    weights <- (2*f_t*S_t)/(1-S_tao^2)

    Cindex1 <- sum(meanRanks * weights) #C-index

    ### Marker 2
    mmm <- MeanRank( survival.time= timeVar, survival.status= eventVar, marker= markerVar2 )

    #Get overlap between survival function and mmm
    meanRanks <-  mmm$mean.rank[which(mmm$time <= cutoffTime)]
    survTimes <- mmm$time[mmm$time <= cutoffTime]
    timeMatch <- match(survTimes, kmfit$time)
    S_t <- kmfit$surv[timeMatch]

    #Calculate weights for c-index
    f_t <- c( 0, (S_t[-length(S_t)] - S_t[-1]) )
    S_tao <- S_t[length(S_t)]
    weights <- (2*f_t*S_t)/(1-S_tao^2)

    Cindex2 <- sum(meanRanks * weights) #C-index

    resultStar[i] <- Cindex1 - Cindex2
  }
  return( quantile(resultStar, probs=c(0.025, 0.975)) )
}
getCindexBstrapCI(nBoot=500, inData=bDat, markerVarName1="score5baseline", markerVarName2="score4baseline",
    timeVarName="time", eventVarName="deathEver", cutoffTime=units*10)
```

```
###FIGURES
#Figure 2
par(mfrow=c(1,2))
par(ps=10)

#AUC I/D
mmmBaseline <- MeanRank(survival.time=bDat$time, survival.status=bDat$deathEver, marker=bDat$score5baseline)
print(length(mmmBaseline$time))
nnnBaseline <- nne(x=mmmBaseline$time, y=mmmBaseline$mean.rank, lambda=0.2, nControls=mmmBaseline$nControls)
plot( mmmBaseline$time, mmmBaseline$mean.rank, xlab="Time (years)", ylab=expression(AUC^"I/D"*"(t)"),
    ylim=c(0.4,1), col="blue", pch=21, cex=.8, axes=F, xlim=c(0,11)*units)
axis(1, at=seq(0,10,by=2)*units, labels=seq(0,10,by=2))
axis(2)
box()
abline(h=0.5, lty=2)
lines( nnnBaseline$x, nnnBaseline$nne, col="blue", lwd=2 )

mmmBaseline <- MeanRank(survival.time=bDat$time, survival.status=bDat$deathEver, marker=bDat$score4baseline)
print(length(mmmBaseline$time))
nnnBaseline <- nne(x=mmmBaseline$time, y=mmmBaseline$mean.rank, lambda=0.2, nControls=mmmBaseline$nControls)
points( mmmBaseline$time, mmmBaseline$mean.rank, col="orange", pch=21, cex=.8 )
lines( nnnBaseline$x, nnnBaseline$nne, col="orange", lwd=2 )

#ROC I/D (TPR)
fpr <- 0.1

mmmBaseline <- dynamicTP( p=fpr, survival.time= bDat$time, survival.status= bDat$deathEver,
    marker= bDat$score5baseline )
print(length(mmmBaseline$time))
nnnBaseline <- nne_TPR(x=mmmBaseline$time, y=mmmBaseline$mean.rank, lambda=0.3,
    nControls=mmmBaseline$nControls, nCases= mmmBaseline$nCases, p=fpr, survival.time= bDat$time,
    survival.status=bDat$deathEver, marker= bDat$score5baseline )  #Fixed bandwidth
plot(mmmBaseline$time, mmmBaseline$mean.rank, xlab="Time (years)",
    ylab=expression(ROC[t]^"I/D"*"(FPF=10%)"), ylim=c(0,1), col="blue", pch=21, cex=.8, axes=F,
```

```r
    xlim=c(0,11)*units)
axis(1, at=seq(0,10,by=2)*units, labels=seq(0,10,by=2))
axis(2)
box()
abline(h=0.5, lty=2)
lines( nnnBaseline$x, nnnBaseline$nne, col="blue", lwd=2 )

mmmBaseline <- dynamicTP(p=fpr, survival.time=bDat$time, survival.status=bDat$deathEver,
    marker= bDat$score4baseline )
print(length(mmmBaseline$time))
nnnBaseline <- nne_TPR( x= mmmBaseline$time, y= mmmBaseline$mean.rank, lambda=0.3,
    nControls= mmmBaseline$nControls, nCases= mmmBaseline$nCases, p=fpr, survival.time= bDat$time,
    survival.status= bDat$deathEver, marker= bDat$score4baseline )  #Fixed bandwidth
points( mmmBaseline$time, mmmBaseline$mean.rank, col="orange", pch=21, cex=.8 )
lines( nnnBaseline$x, nnnBaseline$nne, col="orange", lwd=2 )
legend(x=1.5*units, y=0.9, legend=c("4 covariates", "5 covariates"), col=c("orange","blue"), lty=1, lwd=2,
    horiz=T)

#Figure 3
mmm <- MeanRank(survival.time=pbc2$tstop, survival.status= pbc2$death, marker= currVar, start=pbc2$tstart)

par(mfrow=c(1,2))
par(ps=10)

#AUC I/D
mmmTV <- MeanRank(survival.time=pbc2$tstop, survival.status=pbc2$death, marker=pbc2$score5tv,
    start=pbc2$tstart)
print(length(mmmTV$time))
nnn <- nne( x= mmmTV$time, y= mmmTV$mean.rank, lambda=0.2, nControls=mmmTV$nControls )
plot( mmmTV$time, mmmTV$mean.rank, xlab="Time (years)", ylab=expression(AUC^"I/D"*"(t)"), ylim=c(0.4,1),
    col="blue", pch=21, cex=.8, axes=F, xlim=c(0,11)*units)
axis(1, at=seq(0,10,by=2)*units, labels=seq(0,10,by=2))
axis(2)
box()
abline(h=0.5, lty=2)
```

```r
lines( nnn$x, nnn$nne, col="blue", lwd=2, lty=1 )

mmmTV <- MeanRank(survival.time=pbc2$tstop, survival.status=pbc2$death, marker=pbc2$score4tv,
    start=pbc2$tstart)
print(length(mmmTV$time))
nnn <- nne( x= mmmTV$time, y= mmmTV$mean.rank, lambda=0.2, nControls=mmmTV$nControls )  #Fixed bandwidth
points( mmmTV$time, mmmTV$mean.rank, col="orange", pch=21, cex=.8)
lines( nnn$x, nnn$nne, col="orange", lwd=2, lty=1 )

#ROC I/D (TPR)
mmmTV <- dynamicTP( p=fpr, survival.time =pbc2$tstop, survival.status= pbc2$death, marker= pbc2$score5tv,
 start= pbc2$tstart )
print(length(mmmTV$time))
nnn <- nne_TPR(x=mmmTV$time, y=mmmTV$mean.rank, lambda=0.3, nControls=mmmTV$nControls, nCases=mmmTV$nCases,
    p=fpr, survival.time=pbc2$tstop, survival.status= pbc2$death, marker= pbc2$score5tv, start= pbc2$tstart)
plot( mmmTV$time, mmmTV$mean.rank, xlab="Time (years)", ylab=expression(ROC[t]^"I/D"*"(FPF=10%)"),
    ylim=c(0,1), col="blue", pch=21, cex=.8, axes=F, xlim=c(0,11)*units)
axis(1, at=seq(0,10,by=2)*units, labels=seq(0,10,by=2))
axis(2)
box()
abline(h=0.5, lty=2)
lines( nnn$x, nnn$nne, col="blue", lwd=2 )

mmmTV <- dynamicTP( p=fpr, survival.time =pbc2$tstop, survival.status= pbc2$death, marker= pbc2$score4tv,
    start=pbc2$tstart )
nnn <- nne_TPR(x=mmmTV$time, y=mmmTV$mean.rank, lambda=0.3, nControls=mmmTV$nControls, nCases=mmmTV$nCases,
    p=fpr, survival.time=pbc2$tstop, survival.status=pbc2$death, marker=pbc2$score4tv, start=pbc2$tstart )
points( mmmTV$time, mmmTV$mean.rank, col="orange", pch=21, cex=.8 )
lines( nnn$x, nnn$nne, col="orange", lwd=2 )
legend(x=2.5*units, y=0.3, legend=c("4 covariates", "5 covariates"), col=c("orange","blue"), lty=1, lwd=2,
    horiz=T)

##Figure 4: With CIs for baseline and updated risk scores from 5-covariate model using I/D approach
par(mfrow=c(1,2))
par(ps=10)
```

```r
#ROC I/D
mmmBaseline <- MeanRank(survival.time=bDat$time, survival.status=bDat$deathEver, marker=bDat$score5baseline)
nnnBaseline <- nne(x=mmmBaseline$time, y=mmmBaseline$mean.rank, lambda=0.2, nControls=mmmBaseline$nControls)
plot( mmmBaseline$time, mmmBaseline$mean.rank, xlab="Time (years)", ylab=expression(AUC^"I/D"*"(t)"),
    ylim=c(0.4,1), col="lightblue", pch=21, cex=.8, axes=F, xlim=c(0,11)*units)
axis(1, at=seq(0,10,by=2)*units, labels=seq(0,10,by=2))
axis(2)
box()
abline(h=0.5, lty=2)
lines( nnnBaseline$x, nnnBaseline$nne, col="blue", lwd=2 )
lines( nnnBaseline$x, nnnBaseline$nne + 1.96*sqrt(nnnBaseline$var), col="blue", lty=2 )
lines( nnnBaseline$x, nnnBaseline$nne - 1.96*sqrt(nnnBaseline$var), col="blue", lty=2 )

mmmTV <- MeanRank( survival.time= pbc2$tstop, survival.status= pbc2$death, marker= pbc2$score5tv,
    start= pbc2$tstart )
nnn <- nne( x= mmmTV$time, y= mmmTV$mean.rank, lambda=0.2, nControls=mmmTV$nControls )  #Fixed bandwidth
plot( mmmTV$time, mmmTV$mean.rank, xlab="Time (years)", ylab=expression(AUC^"I/D"*"(t)"), ylim=c(0.4,1),
    col="lightblue", pch=21, cex=.8, axes=F, xlim=c(0,11)*units)
axis(1, at=seq(0,10,by=2)*units, labels=seq(0,10,by=2))
axis(2)
box()
abline(h=0.5, lty=2)
lines( nnn$x, nnn$nne, col="blue", lwd=2, lty=1 )
lines( nnn$x, nnn$nne + 1.96*sqrt(nnn$var), col="blue", lty=2 )
lines( nnn$x, nnn$nne - 1.96*sqrt(nnn$var), col="blue", lty=2 )
```